\documentstyle[12pt]{article}
\def\hybrid{\topmargin 0pt      \oddsidemargin 0pt
	\headheight 0pt \headsep 0pt
	\textheight 9in         
	\textwidth 6.25in       
	\marginparwidth .875in
	\parskip 5pt plus 1pt   \jot = 1.5ex}

\catcode`\@=11
\def\marginnote#1{}
\newcount\hour
\newcount\minute
\newtoks\amorpm
\hour=\time\divide\hour by60
\minute=\time{\multiply\hour by60 \global\advance\minute by-\hour}
\edef\standardtime{{\ifnum\hour<12 \global\amorpm={am}%
	\else\global\amorpm={pm}\advance\hour by-12 \fi
	\ifnum\hour=0 \hour=12 \fi
	\number\hour:\ifnum\minute<10 0\fi\number\minute\the\amorpm}}
\edef\militarytime{\number\hour:\ifnum\minute<10 0\fi\number\minute}

\def\draftlabel#1{{\@bsphack\if@filesw {\let\thepage\relax
   \xdef\@gtempa{\write\@auxout{\string
      \newlabel{#1}{{\@currentlabel}{\thepage}}}}}\@gtempa
   \if@nobreak \ifvmode\nobreak\fi\fi\fi\@esphack}
	\gdef\@eqnlabel{#1}}
\def\@eqnlabel{}
\def\@vacuum{}
\def\draftmarginnote#1{\marginpar{\raggedright\scriptsize\tt#1}}

\def\draft{\oddsidemargin -.5truein
	\def\@oddfoot{\sl preliminary draft \hfil
	\rm\thepage\hfil\sl\today\quad\militarytime}
	\let\@evenfoot\@oddfoot \overfullrule 3pt
	\let\label=\draftlabel
	\let\marginnote=\draftmarginnote
   \def\@eqnnum{(\theequation)\rlap{\kern\marginparsep\tt\@eqnlabel}%
\global\let\@eqnlabel\@vacuum}  }

\def\numberbysection{\@addtoreset{equation}{section}
	\def\theequation{\thesection.\arabic{equation}}}

\def\underline#1{\relax\ifmmode\@@underline#1\else
	$\@@underline{\hbox{#1}}$\relax\fi}

\def\titlepage{\@restonecolfalse\if@twocolumn\@restonecoltrue\onecolumn
     \else \newpage \fi \thispagestyle{empty}\c@page\z@
	\def\thefootnote{\fnsymbol{footnote}} }

\def\endtitlepage{\if@restonecol\twocolumn \else  \fi
	\def\thefootnote{\arabic{footnote}}
	\setcounter{footnote}{0}}  
\catcode`@=12
\relax

\def\beq{\begin{equation}}
\def\eeq{\end{equation}}
\def\bea{\begin{eqnarray}}
\def\eea{\end{eqnarray}}
\def\bar{\overline}

\def\nn{\nonumber}

\relax
\hybrid
\begin{document}
\begin{titlepage}
\setcounter{page}{0}
\begin{center}
 \hfill   PAR--LPTHE 91--52\\
 \hfill         October 1991\\[.4in]
{\large CORRELATION FUNCTIONS OF LOCAL OPERATORS
IN 2D GRAVITY COUPLED TO MINIMAL MATTER\footnote{Talk given at
Carg\`ese Summer School, July 15 - July 27 1991, and at the French -
Russian Workshop, Carg\`ese, July 28 - August 4 1991.}}\\[.4in]
	\large    Vl.S.Dotsenko\footnote{E-mail address:
	dotsenko@lpthe.jussieu.fr}\footnote{Permanent address: Landau
	Institute for Theoretical
Physics, Moscow.}\\[.2in]
	{\it LPTHE\/}\footnote{Laboratoire associ\'e No. 280 au
CNRS}\\
       \it  Universit\'e Pierre et Marie Curie, PARIS VI\\
	Tour 16, 1$^{\it er}$ \'etage \\
	4 place Jussieu\\
	75252 Paris CEDEX 05, FRANCE\\

\end{center}

\vskip .3in
\centerline{ ABSTRACT}
\begin{quotation}
Recent advances are being discussed on the calculation, within the
conformal field theory approach, of the correlation functions for local
operators in the theory of 2D gravity coupled to the
minimal models of matter.

\end{quotation}
\end{titlepage}
\newpage

Here I would like to discuss the results available on the calculation
of the correlation functions of local operators in the theory of
2D gravity coupled to the minimal matter, in the approach of
conformal field theory \cite{pol,cgd,kpz,ddk}.

In the representation of David, Distler and Kawai \cite{ddk}
the local operators take the form:
\beq
\Phi_{s'.s}(z)=\phi^{M}_{s'.s}(z)\phi^{L}_{s'.s}(z)\sim V_{s'.s}(z)
U_{-s'.s}(z)\equiv \exp(i\alpha_{s'.s}\varphi^{M}(z))
\exp(\beta_{-s'.s}\varphi^{L}(z)) \label{L1}
\eeq
where
\bea
\alpha_{s'.s}=\frac{1-s'}{2}\alpha_{-}+\frac{1-s}{2}\alpha_{+},
\quad \beta_{-s'.s}=\frac{1+s'}{2}\beta_{-}+\frac{1-s}{2}\beta_{+}
\label{L2}\\
\alpha_{\pm}=\alpha_{0}\pm \sqrt{\alpha^{2}_{0}+2},
\quad\beta_{\pm}=\beta_{0}\pm \sqrt{\beta^{2}_{0}-2} \label{L3}
\eea
Here the free field representation is assumed both for the matter
and for the gravity (Liouville) factors of the operator (\ref{L1}).
More specifically:
\bea
\langle \varphi^{M}(z)\varphi^{M}(z')\rangle=
\langle \varphi^{L}(z)\varphi^{L}(z')\rangle=\log \frac{1}{z-z'}
\label{L4}\\
T^{M}=-\frac{1}{2} \partial \varphi^{M} \partial \varphi^{M}
+i \alpha_{0} \partial^{2} \varphi^{M}, \quad\ T^{L}
=-\frac{1}{2}\partial\varphi^{L}\partial\varphi^{L}+\beta_{0}
\partial^{2}\varphi^{L} \label{L5}\\
C_{M}=1-12\alpha^{2}_{0},\quad C_{L}=1+12\beta^{2}_{0} \label{L6}\\
V_{\alpha}=\exp(i\alpha\varphi^{M}),\quad
U_{\beta}=\exp(\beta\varphi^{L}) \label{L7}\\
\bigtriangleup^{M}_{\alpha}=\frac{1}{2}(\alpha^{2}-2\alpha\alpha_{0}),
\quad
\bigtriangleup^{L}_{\beta}=-\frac{1}{2}(\beta^{2}-2\beta\beta_{0})
\label{L8}
\eea
The constraint which couples the two theories is
\beq
C_{M}+C_{L}=26 \label{L9}
\eeq
and the constraint which couples the two representations of the
corresponding Virasoro algebras is
\beq
\bigtriangleup^{M}+\bigtriangleup^{L}=1 \label{L10}
\eeq
Eq.(\ref{L9}) leads to
\beq
\beta^{2}_{0}=\alpha^{2}_{0}+2,\quad \beta_{\pm}=\pm\alpha_{\pm}
\label{L11}
\eeq
while the expressions in (\ref{L2}) for $\alpha, \beta$ solve
for (\ref{L10}), on account of eqs.(\ref{L8}) for
$\bigtriangleup^{M},\bigtriangleup^{L}$.

The notations for the Liouville part, made to be symmetric
to the matter sector, are those used in \cite{d}.

The three-point functions, or amplitudes, take the form
\bea
A^{(3)} &=& \langle \Phi_{s'.s}(0) \Phi_{n'.n}(1)
      \Phi_{m'.m}(\infty)\rangle \nn\\
	&=& \langle \phi^{M}_{s'.s}(0) \phi^{M}_{n'.n}(1)
\phi^{M}_{m'.m}(\infty) \rangle
\langle \phi^{L}_{s'.s}(0) \phi^{L}_{n'.n}(1)
\phi^{L}_{m'.m}(\infty)\rangle=
A^{(3)}_{M} A^{(3)}_{L} \label{L12}
\eea
The three-point functions of the matter sector had been calculated in
\cite{df}, and can be given in the following form:
\bea
A^{(3)}_{M} &\propto& (\pi\gamma(\rho'))^{l}(\pi\gamma(\rho))^{k}
 P(l,k)P(l-s',k-s)P(l-n',k-n)P(l-m',k-m) \nn\\
   &\times& P^{-1}(l-s',k-s)P^{-1}(l-n',k-n)P^{-1}(l-m',k-m)
   \label{L13}
\eea
Here
\bea
P(l,k) &=& \prod^{l}_{i=1} \frac{\Gamma(i\rho')}{\Gamma(1-i\rho')}
       \prod^{k}_{j=1} \frac{\Gamma(j\rho)}{\Gamma(1-j\rho)}
       \prod^{l}_{i=1} \prod^{k}_{j=1}\frac{(-1)}{(i\rho'-j)^{2}}
       \label{L14}\\
 \rho &=& \frac{\alpha^{2}_{+}}{2}, \quad
 \rho'=\frac{\alpha^{2}_{-}}{2}
 =\frac{1}{\rho},\quad \gamma(\rho)=\frac{\Gamma(1-\rho)}{\Gamma(\rho)}
\label{L15}\\
 l &=& \frac{s'+n'+m'-1}{2}, \quad  k=\frac{s+n+m-1}{2} \label{L16}
\eea
The expression in (\ref{L15}) is different from that in \cite{df}
by normalization factors of individual operators, and is the same as
in \cite{d}. It corresponds to the Coulomb gas operators without
extra normalization, apart from possible sign and $\rho^{(...)}$
factors.

The expression for $A^{(3)}_{L}$ in (\ref{L12}) can be obtained by
an analytic continuation of the result for $A^{(3)}_{M}$ (\ref{L13})
on using the following analytic continuation for the finite products
with the negative integer upper bounds \cite{d} :
\beq
\hbox{if} \quad q(l)=\prod^{l}_{i=1} f(i), \quad \hbox{then} \quad
q(l=-|l|)=\prod^{|l|-1}_{i=0} \frac{1}{f(-i)} \label{L17}
\eeq
Multiplying $A^{(3)}_{M}$ and $A^{(3)}_{L}$, eq.(\ref{L12}),
one gets \cite{d} :
\beq
A^{(3)} \propto (\mu)^{S} N_{s'.s} N_{n'.n} N_{m'.m} \label{L18}
\eeq
Here
\bea
N_{s'.s}=\frac{\Gamma(1+s'\rho'-s)}{\Gamma(-s'\rho'+s)}
&=&\frac{\Gamma(\frac{\beta^{2}-\alpha^{2}}{2})}{\Gamma(1
-\frac{\beta^{2}-\alpha^{2}}{2})}, \quad \alpha=\alpha_{s'.s},
\quad \beta=\beta_{-s'.s} \label{L19}\\
S&=&\frac{1}{\beta_{-}}(2\beta_{0}-\beta_{-s'.s}-\beta_{-n'.n}
-\beta_{-m'.m}) \label{L20}
\eea
The three-point functions of the minimal model coupled to gravity
first had been calculated in \cite{dk}, using the KdV technique
of \cite{doug}. In \cite{gl} they had been obtained in the
Liouville theory approach, by integrating out first the Liouville
field zero mode, the trick introduced in \cite{gtw}, and than by
using the analytic continuation technique,
though different from the one sketched on above.

Both in \cite{dk} and \cite{gl} the calculation had been restricted
to the order operators, i.e. to the operators with
\beq
s'=s,\quad n'=n,\quad m'=m \label{L21}
\eeq
The three-point functions for the general operators
had been calculated in \cite{d}, with result (\ref{L18}).
We remark that general operators are known to be quite a problem for
the KdV
technique, and may be this is the first instance when the field
theory happens to be more powerful with respect to the otherwise
quite successful matrix model approach \cite{bk} and the related
KdV technique \cite{doug,dk}.

We shall now discuss some of the properties of the three-point
functions.

1. Invariance to switching the Fock space representations.

Each operator factor in (\ref{L1}) has two possible representations
in the free field, or the Coulomb gas technique:
\bea
\phi^{M} \hbox{:} \quad V_{s'.s}, \quad V_{-s'.-s} \label{L22}\\
\phi^{L} \hbox{:} \quad U_{-s'.s}, \quad U_{s'.-s} \label{L23}
\eea
So we can use four representations:
\bea
\Phi^{--}_{s'.s} \sim V_{s'.s}U_{-s'.s} \label{L24}\\
\Phi^{-+}_{s'.s} \sim V_{s'.s}U_{s'.-s} \label{L25}
\eea
and two more, obtained with $V_{s'.s} \rightarrow V_{-s'.-s}$.
The tree-point functions can be calculated for these different
representations, with the analytic continuation technique
sketched on above, to be used also for the matter sector,
with the result as in (\ref{L18}) but with the normalization factors
replaced according to:
\bea
\Phi^{--}_{s'.s} \sim N^{--}_{s'.s}=\frac{\Gamma(1
 +s'\rho'-s)}{\Gamma(-s'\rho'+s)} \label{L26}\\
\Phi^{-+}_{s'.s} \sim N^{-+}_{s'.s}=\frac{\Gamma(1
 -s'+s\rho)}{\Gamma(s'-s\rho)} \label{L27}
\eea
with two more obtained by switching signs. (See also the
calculations in\cite{agbg}). The form
\beq
N=\frac{\Gamma(\frac{\beta^{2}-\alpha^{2}}{2})}{\Gamma(1
-\frac{\beta^{2}-\alpha^{2}}{2})} \label{L28}
\eeq
for $\Phi=V_{\alpha}U_{\beta}$ always holds. In this sense the
three-point functions are invariant with respect to switching the
representations. This is the case also for the matter theory, i.e.
the minimal model itself. The functions
\beq
\langle \phi_{i} \phi_{j} \phi_{k} \rangle,
\quad \langle \phi_{i} \phi_{j}\tilde{ \phi}_{k} \rangle,
\quad \langle \phi_{i} \tilde{\phi}_{j}\tilde{ \phi}_{k} \rangle,
\quad \langle \tilde{\phi}_{i} \tilde{\phi}_{j}\tilde{ \phi}_{k}
\rangle,\label{L29}
\eeq
($\phi_{i}\sim V_{s'.s},\quad \tilde{\phi}_{i}\sim V_{-s'.-s}$)
differ by normalization factors of individual operators.

It appears reasonable to expect that for more-point functions
there should exist some form of equivalence of the representations,
if proper formulation is found. Since, whatever the representation,
we are dealing with two coupled conformal theories, not with
particular Fock spaces. Again, this is the case for the matter theory.
E.g. for the four-point functions one usually uses the
representation $\langle \phi\phi\phi \tilde{\phi} \rangle $
\cite{df}. In fact, the representation
$\langle \tilde{\phi} \tilde{\phi} \tilde{\phi}\phi
\rangle $ could be used equally well, by employing antiscreening
operators (or, a negative number of screening operators), as
intertwiners of the Fock spaces.
This extension of the
technique would restore the direct - conjugate representation
symmetry. By the way, the antiscreening operators could be useful
objects from the point of view of the quantum group representation.

Returning to gravity, we remark still that the direct - conjugate
representation symmetry had already been assumed in the form of the
effective action for the quantum theory of Liouville \cite{d}:
\beq
A[\phi^{L}] \propto \int [\partial \phi \bar{\partial} \phi
+\mu \exp(\beta_{-}\phi^{L})+a(\mu)^{\rho}\exp(\beta_{+}\phi^{L})]
\label{L30}
\eeq
which is the free field representation with two screening operators,
more proper to say. (Some arguments to such a
representation had been given also in \cite{dhok}).

In (\ref{L30})
\beq
\mu \exp(\beta_{-}\phi^{L})=\mu \Phi^{--}_{1.1}=\mu V_{1.1}U_{-1.1}
\label{L31}
\eeq
is the puncture operator, and
\beq
(\mu)^{\rho}\exp(\beta_{+}\phi^{L})=(\mu)^{\rho}\Phi^{-+}_{1.1}
=(\mu)^{\rho}V_{1.1}U_{1.-1}\label{L32}
\eeq
is the L-sector conjugate representation operator. Then it should be
natural to assume the possibility of using, on equal footing,
the alternative representations for other operators as well, being
multiplied by the $\mu$-scaling compensating factors.

2. The result (\ref{L18}) assumes that the usual fusion rules of the
minimal model are cancelled by gravity. In particular, the operators
outside the basic conformal grid couple to the states inside, and so
they become physical.
This result is obtained also by the analytic continuation technique
of \cite{gl}, extended to more general set of operators in \cite{kit}.
This disagrees with the KdV results of \cite{dk}, for the order
operators, which support instead the usual fusion rules of the minimal
model. On the other hand, coupling of operators outside the basic
conformal grid may be in agreement with the BRST analysis of the paper
\cite{lz}, for the physical states (or operators) spectrum.
This has been further investigated recently in \cite{muk,bouwk}.
Although the extra states found in \cite{lz} (for the case of minimal
$C<1$ models coupled to gravity) involve ghosts, further analysis of
\cite{bouwk} assumes that they could equivalently be represented
by the states (operators) outside the basic conformal grid, without
ghosts, with their ghost number grading replaced by the Felder BRST
grading \cite{fel} (the number of block steps away from the basic
conformal
grid).

The conflicting evidence stated above, due to different techniques
employed, shows that the problem with the fusion rules is still open.
See also the discussions in \cite{agbg,agg}. We expect that, as in
the case of the minimal conformal theory itself,
to fix the three-point functions we have to understand the
four-point ones. They involve a lot more dynamics. We shall describe
next some advances made in the calculation of four-point functions,
in the field theory approach.

The Coulomb gas representation for the four-point functions has the
form (we drop the $\mu$ scaling factor throughout):
\bea
&A^{(4)}& = \int d^{2}z \langle \Phi(0) \Phi(z) \Phi(1) \Phi(\infty)
\rangle\nn\\
&=&\int d^{2}z [(\prod^{l}_{i=1} \int d^{2}x_{i})
(\prod^{k}_{j=1} \int d^{2}y_{j})
\langle
V_{\alpha 1}(0)V_{\alpha 2}(z)V_{\alpha 3}(1)V_{\alpha 4}(\infty)
(\prod^{l}_{i=1} V_{-}(x_{i}))
(\prod^{k}_{j=1} V_{+}(y_{j})) \rangle \nn\\
&\times&(\prod^{\tilde{l}}_{i=1} \int d^{2}u_{i})
(\prod^{\tilde{k}}_{j=1} \int d^{2}v_{j})
\langle
U_{\beta 1}(0) U_{\beta 2}(z) U_{\beta 3}(1) U_{\beta 4}(\infty)
(\prod^{\tilde{l}}_{i=1} U_{-}(u_{i}))(\prod^{\tilde{k}}_{j=1}
U_{+}(v_{j}))
\rangle]\label{L33}
\eea
The matter and the Liouville correlation functions in (\ref{L33})
have separate expansions over the $z$ singularities,
$(z)^{-2(\bigtriangleup^{M}_{1}+\bigtriangleup^{M}_{2}
+\bigtriangleup^{M}_{int})},\quad
(z)^{-2(\bigtriangleup^{L}_{1}+\bigtriangleup^{L}_{2}
+\bigtriangleup^{L}_{int})}$, which correspond to the intermediate
primary operators (states). When multiplied, they produce the double
expansion of the form:
\beq
\sum_{\tilde{p'},\tilde{p}}\sum_{p',p}\int d^{2}z
 |z|^{-2(\bigtriangleup_{1}+\bigtriangleup_{2}
+\bigtriangleup_{int})}\times  (...)\label{L34}
\eeq
where
\beq
\bigtriangleup_{1}=\bigtriangleup^{M}_{1}+\bigtriangleup^{L}_{1}=1,
\quad
\bigtriangleup_{2}=\bigtriangleup^{M}_{2}+\bigtriangleup^{L}_{2}=1,
\quad
\bigtriangleup_{int}=\bigtriangleup^{M}_{p'.p}
+\bigtriangleup^{L}_{\tilde{p'}.\tilde{p}} \label{L35}
\eeq
The terms in (\ref{L34}) with $\tilde{p'}=-p',\tilde{p}=p$
(the `diagonal' ones), such that
$\bigtriangleup_{int}
=\bigtriangleup^{M}_{p'.p}+\bigtriangleup^{L}_{-p'.p}=1$,
they correspond to physical intermediate states. The integral
(\ref{L33}) then gets log singularities, $\int d^{2} |z|^{-2}(...)$.

To better define the integral (\ref{L33}) let us shift the values
of the parameters (Coulomb gas charges, or states momenta) of the
external states, slightly off their discrete values, but so that
the physical states condition is kept, $\bigtriangleup_{\alpha}
+\bigtriangleup_{\beta}=1$, which requires $\beta_{i}=\alpha_{i}
-\alpha_{-}$ (or $\beta_{i}=\alpha_{+}-\alpha_{i}$, if conjugate
representation is used):
\beq
\alpha_{i}=\alpha^{(0)}_{i}+\frac{\epsilon_{i}}{2}\alpha_{-},
\quad
\beta_{i}=\beta^{(0)}_{i}-\frac{\epsilon_{i}}{2}\beta_{-},
\quad
\sum \epsilon_{i}=0 \label{L36}
\eeq
Here
\bea
\alpha^{(0)}_{1}&=&\alpha_{s'.s},\quad
\alpha^{(0)}_{2}=\alpha_{n'.n},\quad
\alpha^{(0)}_{3}=\alpha_{m'.m},\quad
\alpha^{(0)}_{4}=\alpha_{t'.t} \nn\\
\beta^{(0)}_{1}&=&\beta_{-s'.s},\quad \hbox{so on} \label{L37}
\eea
Then one checks that the integral (\ref{L33}) has the following
expansion in $z$ singularities:
\bea
A^{(4)} &\sim& \sum_{k_{1},l_{1}}\sum_{\tilde{k}_{1},\tilde{l}_{1}}
\int d^{2}z
|z|^{-2\gamma_{12}} \times Res \label{L38}\\
\gamma_{12} &=&
1+[(\tilde{l}_{1}+l_{1}+1)+(\tilde{k}_{1}-k_{1})\rho]\nn\\
 &\times&  [(s'+n'+\tilde{l}_{1}-l_{1})\rho'
+(1-s-n+\tilde{k}_{1}+k_{1})-(\epsilon_{1}+\epsilon_{2})\rho']
\label{L39}
\eea
$Res$ in (\ref{L38}) stands for the residue amplitude;
$\tilde{l}_{1},\tilde{k}_{1},l_{1},k_{1}$ are numbers of screening
operators integraded over the region close to $0$ and $z$, and
it is presumed that $z$ is close to $0$.
The rest of screenings, $\tilde{l}_{2}=\tilde{l}-\tilde{l}_{1}$,
$\tilde{k}_{2}=\tilde{k}-\tilde{k}_{1}$,
$l_{2}=l-l_{1}$, $k_{2}=k-k_{1}$ are being away from $z \sim 0$.
(The corresponding analysis in case of the minimal conformal theory
of selecting the $z$-singularities
of four-point functions by the configurations of the screening
operators see in \cite{df}, and also in \cite{dot}). By summing
over the configurations if screening operators in (\ref{L38})
we are summing over the intermediate states in (\ref{L34}).
The terms in (\ref{L38}) with
\beq
\tilde{l}_{1}=-s'-n'+l_{1},\quad \tilde{k}_{1}=s+n-1-k_{1} \label{L40}
\eeq
correspond to physical states. We get in this case
\bea
\gamma_{12}&=&1+(\epsilon_{1}+\epsilon_{2})\rho'(p'-p\rho)\label{L41}\\
p'&=&s'+n'-1-2l_{1},\quad p=s+n-1-2k_{1} \label{L42}
\eea
The subsum over the physical intermediate states, the `diagonal' terms
in (\ref{L38}), which produce $1/\epsilon$ singularities, takes the
form:
\beq
(A^{(4)})_{sing} \sim \sum_{k_{1},l_{1}(p',p)} \int d^{2}z
|z|^{-2-2\rho'(\epsilon_{1}+\epsilon_{2})(p'-p\rho)} \times Res
\label{L43}
\eeq
It is not difficult to realize that the sum for $(A^{(4)})_{sing}$
can be given in the form:
\beq
(A^{(4)})_{sing} \sim \sum_{p',p} \frac{1}{(\epsilon_{1}+\epsilon_{2})
(p'-p\rho)} \langle\Phi^{--}_{s'.s}\Phi^{--}_{n'.n}\Phi^{-+}_{p'.p}
\rangle \langle\Phi^{+-}_{p'.p}\Phi^{--}_{m'.m}\Phi^{--}_{t'.t}
\rangle \label{L44}
\eeq
Using the expressions for the three-point functions
we obtain, for the residue in (\ref{L44}),
\beq
\sum_{p',p}\frac{1}{(p'-p\rho)}N_{1}N_{2}N^{-+}_{p'.p}
N^{+-}_{p'.p}N_{3}N_{4} \label{L45}
\eeq
$N_{1}=N^{--}_{s'.s}$, so on. As
\beq
N^{-+}_{p'.p}N^{+-}_{p'.p}=\frac{\Gamma(1-p'+p\rho)}{\Gamma(p'-p\rho)}
\frac{\Gamma(1+p'-p\rho)}{\Gamma(-p'+p\rho)}=(-1)(p'-p\rho)^{2}
\label{L46}
\eeq
we obtain finally
\beq
(A^{(4)})_{sing} \sim N_{1}N_{2}N_{3}N_{4} \sum_{p',p} (p'-p\rho)
\label{L47}
\eeq

We have calculated the contribution of physical intermediate states
only, and also just the singular piece, the residue at $1/\epsilon$
in $A^{(4)}$, coming from the region $z\sim 0$. Also,
different representations for the external states (operators)
can be used, but the
general form of $(A^{(4)})_{sing}$ remains as in (\ref{L47}).

We notice that although we have chosen the representation
$(--)$ for the external states, see (\ref{L37}),(\ref{L44})
and the difinitions (\ref{L24}),(\ref{L25}), the opposite
representation (or the opposite Liouville dressing)
states appear anyway, as intermediate states
in the four-point functions, see (\ref{L44}). (Same is true
of course for the often prefered representation,
$\langle (--)(--)(--)(--)(+-) \rangle$ , with one of the external
operators
chosen to be in the conjugate matter representation). This is another
piece to the arguments given above on should be equal footing of the
representations.

Several further comments are in order.
We have calculated, in the $\epsilon$
- regularization, the residue at $\epsilon$ - singularity coming
from the region $z \sim 0$ ($s$ channel, in terminology of dual
models) of the integral (\ref{L33}). Similar contributions
should be considered also coming from $z \sim 1$ and $z \sim \infty$
($t$ and $u$ channels). And they have to be summed over, with the
coefficients which are not defined within the $\epsilon$ -
regularization used above. In fact, the regularization fixes
the calculation just for one
particular channel, and leaves
the relative coefficients of different channels not defined.

Related flaw is that the intermediate state violates the relation
$\bigtriangleup^{M}+\bigtriangleup^{L}=1$, by $\epsilon$ amount,
which we kept for the external states.
In this respect the $\epsilon$ - regularization is not fully
consistent.

Other questions have been left open in the above calculation. We
assumed that the correlation functions are given by the residues of
the physical state singularities. What happens to the nonphysical
intermediate states in the full sum (\ref{L38})? One suggestion is
that these problems can be resolved and a fully consistent
regularization can be achieved via the analytic decomposition
of the integral in (\ref{L33}) into a sum of products of contour
integrals. This is the alternative way in which the four-point
(and multipoint) functions can be defined, and the operator algebra
calculated,
in the usual conformal theory, see \cite{df,dot}.
The expectation is that proceeding in this way the contribution of
nonphysical channels will get cancelled, presumably because
there are pairs of intermediate states, with exponents
\beq
\bigtriangleup^{M}+\bigtriangleup^{L}, \quad \hbox{and} \quad
(1-\bigtriangleup^{M})+(1-\bigtriangleup^{L})=2-\bigtriangleup^{M}
-\bigtriangleup^{L} \label{L48}
\eeq
which enter under $\sin\pi(...)$ in the coefficients at products
of the corresponding conformal blocks. The two terms corresponding
to (\ref{L48}) would cancel each other.

Also, in case of terms corresponding to physical states, it is likely
that poles will contribute only ($z$ going around $0$, $1$, $\infty$),
as exponents are integer in this case.

Proceeding in this way it might be possible to actually check
what form of factorization of the four-point functions is
taking place.

One more technical question is the sum in (\ref{L38}), and in
(\ref{L47}), over the numbers of the screening operators.
By their origin, they have to be summed in the ranges:
\beq
0 \leq \tilde{l}_{1} \leq \tilde{l}, \quad
0 \leq \tilde{k}_{1} \leq \tilde{k}, \quad
0 \leq l_{1} \leq l, \quad 0 \leq k_{1} \leq k \label{49}
\eeq
where, by the charge conservation, $\sum \alpha_{i}=2\alpha_{0}$,
$\sum \beta_{i}=2\beta_{0}$, the total numbers of the screenings
are given by:
\bea
l&=&\frac{s'+n'+m'+t'-2}{2},\quad \tilde{l}=\frac{-s'-n'-m'-t'-2}{2}
= -l-2 \label{L50}\\
k&=&\frac{s+n+m+t-2}{2}=\tilde{k} \label{L51}
\eea
So we have to define a sum over a negative number of screenings - over
$l_{1}$, in this particular representation of the external operators.
A simple conjecture would be, by extending the analytic continuation
trick for finite products, eq.(\ref{L17}), that the sums are to be
defined
as:
\beq
\hbox{if} \quad s(l)=\sum^{l}_{i=0} f(i), \quad \hbox{then} \quad
s(l=-|l|)=-\sum^{|l|-1}_{i=-1} f(-i) \label{L52}
\eeq
(l - integer).

For the moment the above remarks are only conjectures and further
work is needed.

We mention that special multipoint functions, which involve
no matter screenings, have been calculated in \cite{dk2}, in the
Liuoville field theory approach, using the
representation of Goulian and Li.
Other special four-point functions could be obtained
by just differentiating the general three-point functions
(\ref{L18}) with respect to $\mu$. These are the four-point
functions which involve one puncture operator.
Those are special
isolated cases by which the general technique could
in particular be verified.

\noindent{\large Acknowledgements}

I am grateful to the colleagues at LPTHE for their kind
hospitality. The efforts and hospitality of the
organizers of the Carg\`ese Summer School, and of the French - Russian
Workshop are also
gratefully acknowledged. The discussions with L. Alvarez-Gaum\'e,
P. Bouwknegt, G. Felder, C. G\'omes, S. Mukhi, and especially with
V. A. Fateev, were both useful and stimulating.


\end{document}